\documentclass[namedreferences, nointegrals]{solarphysics}
\usepackage[hyperref,optionalrh,solaromanenum]{spr-sola-addons}
\usepackage{amsmath,amssymb}
\usepackage{epsfig}
\usepackage{graphicx}
\usepackage{courier}
\usepackage{color}
\usepackage{breakurl}
\usepackage{dcolumn}
\usepackage{multirow}
\usepackage{bm}

\renewcommand{\div}{\operatorname{div}}
\renewcommand{\vec}[1]{{\mathbfit #1}}

\begin{document}
\begin{article}

\begin{opening}

\title{Migrating Dynamo Waves and Consequences for Stellar Current Sheets}

\author[addressref={aff1,aff2},email={emaevskiy@mail.ru}]{\inits{E.}\fnm{Evgeniy}~\lnm{Maiewski}\orcid{0000-0002-9647-3616}}
\author[addressref={aff1,aff3},email={hmalova@yandex.ru}]{\inits{H.}\fnm{Helmi}~\lnm{Malova}\orcid{0000-0001-6511-2335}}
\author[addressref={aff1,aff2,aff4},email={masterlu@mail.ru}]{\inits{V.}\fnm{Victor}~\lnm{Popov}}
\author[addressref={aff4,aff5,aff6},email={sokoloff.dd@gmail.com}]{\inits{D.}\fnm{Dmitry}~\lnm{Sokoloff}\orcid{0000-0002-3441-0863}}
\author[addressref={aff1,aff4,aff6},corref,email={yushkov.msu@mail.ru}]{\inits{E.}\fnm{Egor}~\lnm{Yushkov}\orcid{0000-0002-9810-1541}}

\address[id=aff1]{Space Research Institute of the Russian Academy of Sciences (IKI), 84/32 Profsoyuznaya Str, Moscow, Russia, 117997}
\address[id=aff2]{HSE University (Higher School of Economics), 20 Myasnitskaya Str, Moscow, Russia, 101000}
\address[id=aff3]{Skobeltsyn Institute of Nuclear Physics (SINP MSU), 1(2) Leninskie gory, GSP-1, Moscow, Russia, 119991}
\address[id=aff4]{Physics Department of Moscow State University, 1(2) Leninskie gory, GSP-1, Moscow, Russia, 119991}
\address[id=aff5]{Pushkov Institute of Terrestrial Magnetism, Ionosphere and Radio Wave Propagation Russian Academy of Sciences (IZMIRAN), 4 Kaluzhskoe shosse, Troitsk, Russia, 142190.}
\address[id=aff6]{Moscow Center of Fundamental and Applied Mathematics, 1(2) Leninskie gory, GSP-1, Moscow, Russia, 119991}

\runningauthor{E. Maiewski et al.}
\runningtitle{Migrating Dynamo Waves and Consequences for Stellar Current Sheets}

\begin{abstract}
We study the relation between stellar dynamo-wave propagation and the structure of the stellar magnetic field. Modeling dynamo waves by the well-known Parker migratory dynamo, we vary the intensity of dynamo drivers in order to obtain activity-wave propagation toward the Equator (as in the solar-activity cycle) or towards the Poles. We match the magnetic field in the dynamo active shell with that in the surrounding stellar material, using a simple dissipativ magnetohydrodynamic system for the transition region. Introducing a weak asymmetry between the stellar hemispheres, we study phase shifts of the dipole, quadrupole, and octupole magnetic components at various distances from the star to demonstrate that several-percent asymmetry in dynamo drivers are sufficient to obtain a realistic relation between solar dipole and quadrupole moments. We study the behavior of the stellar current sheets and show that for the poleward propagating activity it is substantially different from solar ones. In particular, we demonstrate conditions in which the conical current sheets propagate opposite to the solar directions.
\end{abstract}

\end{opening}

\section{Introduction}
Investigation of planetary and stellar current sheets is an interesting topic in modern cosmic studies. Space missions have revealed that most of the planets in the solar system, as well as many discovered stars and exoplanets, have their proper magnetic fields \cite[e.g.][]{2014Sci...346..981K,2014SSRv..182...85J,2016ApJ...818...24K, 2019oeps.book...31C,Zetal19, Cetal19,2000eaa..bookE2322B}. Planetary fields have the shapes of cavities (named magnetospheres) in the fast flows of the solar/stellar wind \cite[e.g.][]{1961CaJPh..39.1433A,2019ARep...63..550Z,2000eaa..bookE2322B}. The planetary magnetospheres are usually supported by the complex system of currents flowing on the magnetopause surfaces and closing on magnetotails, i.e. elongated magnetic configurations on the nightside of planets \cite[e.g.][]{1978SSRv...21..489A,Zetal19,2000eaa..bookE2322B}. Large-scale current sheets having disk-like or conic shapes  \citep{2001JGR...10615819S} are observed throughout the whole heliosphere up to the heliopause, where the influence of the solar wind weakens. It is assumed that current sheets in both the heliosphere and stellar astrospheres should have common mechanisms of formation, particularly, associated with the presence of the magnetic fields generated by the interior of stars due to the dynamo processes.

After the discussion of \cite{1961ApJ...134...20P} about the Sun as a non-equilibrium celestial object ejecting plasma with a frozen magnetic field into the surrounding space, almost two decades passed before the relation was established between the cross-section of the interplanetary magnetic field (IMF) sector boundaries by the Earth and the existence of a corresponding large-scale current sheet in the heliosphere. Thus \cite{1976SvalgaardNatur.262..766S} described the general structure of the IMF and interpreted the intersections of magnetic sectors as cross-sections of the folded current sheet, where the tangential component of the IMF has opposite signs on each of its sides \citep{1980RosenbergJGR....85.3021R}. Later, on the basis of spacecraft data, the nature of the heliospheric current sheet was revealed as an extension of the streamer belt \citep{2005RobertsJGRA..110.6102R} that is visible during magnetically quiet periods in the solar corona as a chain of helmet-like magnetic structures along the Equator \citep{1995WooSSRv...72..223W, 1996CrookerJGR...10124331C, 1999SoPh..188..277E} separating regions with opposite magnetic fluxes. 

The heliospheric current sheet that is formed at some distance from the Sun appeared to be an extremely thin disk-like structure with thickness about 10,000 km embedded inside a much thicker (about tens of solar radii) heliospheric plasma sheet \citep{1997BavassanoGeoRL..24.1655B}. However, it remained unclear for a long time what happens with the heliospheric current sheet during active periods of solar activity, when the dipole and multipole harmonics become strong and competitive, and as a result the streamer belt can have a complicated shape. It was shown by \cite{1996WangApJ...456L.119W} that at that some periods multipole and dipole harmonics are mixed and as a result the current sheet can be deflected to higher latitudes and can have a strongly folded shape \citep{2014WangSSRv..186..387W}. Similar theoretical conclusions were confirmed later by \cite{2001JGR...10615819S}. As was shown by \cite{2004MursulaSoPh..224..133M} the multipole expansion in the solar corona has a strong quadrupole term, which is oppositely directed to the dipole one. These results implied that the Sun has a symmetric quadrupole dynamo mode that oscillates in phase with the dominant dipole mode. Moreover, the heliospheric magnetic field has a strong tendency to produce solar tilts that are roughly opposite in longitudinal phase. It was concluded that the solar dynamo includes three modes: quadrupole, dipole, and non-axisymmetric ones. 

Now most scientists agree that the heliospheric current sheet is formed as an extension of the streamer belt in the solar corona \citep[e.g.][]{1999SoPh..188..277E, 2005RobertsJGRA..110.6102R}. Thus the current sheet surface in the heliosphere generally corresponds to the geometry of the neutral line of the Sun. In particular, during minima of solar activity, when the geomagnetic field is dipole and the neutral line is located along the solar equator, the heliospheric current sheet is observed at low latitudes and has the known disk-like shape \citep{2001JGR...10615819S}. The theoretical consideration shows that during periods of maximum solar activity, when the quadrupole or octupole magnetic components compete with the dipole one, the shape of the neutral line (or lines) can be very complex; for example, there may be two or three large-scale current sheets in the heliosphere \citep{2015ApJ...798..116R, kislov2019quasi, maiewski2020magnetohydrodynamic}, so that the conic-like sheet forms at high latitudes \citep{2017KhabarovaApJ...836..108K}, while at low latitudes the current sheet tends to be disk-like \citep{2001JGR...10615819S, kislov2019quasi, maiewski2020magnetohydrodynamic}, although this has not yet been proven observationally.

It should be mentioned that the relation of the structure and dynamics of solar/stellar magnetic fields with the dynamo processes inside them has not been studied well. The most consistent earlier publications were presented recently by \cite{ maiewski2020magnetohydrodynamic} and \cite{sym12122085}. While these results were investigated mostly in theory, in practice the Zeeman--Doppler imaging studies appeared as a useful and available tools to investigate the structure and evolution of stellar magnetic fields \citep{1958ApJS....3..141B, 2015AJ....150..146B, 2015SSRv..191...27L}. Particularly, such studies revealed the peculiarities of the large-scale magnetic fields of accreting pre-main-sequence stars \citep{2011AmJPh..79..461G}. It was shown that rotating stars can possess general magnetic fields in the range from several Gs to tens of kGs evolving in time \citep[e.g.][and references therein]{1958ApJS....3..141B, 2021A&ARv..29....1K}. The magnetic fields of some stars can periodically change their strength and direction of large-scale magnetic fluxes \citep{1949PASP...61..226B}. For other stars the strengths of a large-scale magnetic field can strongly fluctuate, but the changes of directions of magnetic fluxes do not occur \citep{1956ApJ...124..489B}. For example, the star HD $188041$ was found to have magnetic field of a stable polarity with magnetic strength variation in the diapason from 600 to 4800 G in a period about 226 days \citep{1954ApJ...120...66B}. However, the shapes of large-scale stellar magnetic fields can demonstrate a quite large variety \citep{1961ApJ...134...20P}. Thus the main-sequence stars and compact objects such as white dwarfs or neutron stars have magnetic fields that strongly deviate from a pure dipole located exactly in the star’s center. These magnetic fields can be approximated by an off-centered dipole or multipole fields consisting from quadrupole or general lowest-order multipole combinations \citep[e.g.][]{2019ApJ...887L..23B, 2020MNRAS.499.4445P, de2020evidence}. On the other hand, the pre-main-sequence stars all have multipolar magnetic fields with a dominating octupole component \citep{2011AmJPh..79..461G}

A significant contribution to the study of the relationship between dynamo processes, stellar wind, and IMF was made in Pinto's numerical experiments \citep[see, e.g.,][]{pinto2011coupling, pinto2013coupling, pinto2016flux}. In these articles, the results of 2.5-dimensional axially symmetric kinematic models (STELEM) and 2.5-dimensional isothermal MHD code (DIP) were stitched together and the changes in stellar-wind properties, momentum, and mass losses at different stages of the solar cycle were studied. Comparison of these numerical results with PSP data demonstrated the realism of the stitching of magnetic structures, as well as the fact that, in general terms, we understand the essence of the process \citep{telloni2021exploring}. Similar numerical experiments on magnetic-field ejection from the Sun using the PENCIL code were carried out by \cite{warnecke2011dynamo} and \cite{warnecke2014cause}, where they clearly demonstrated the essential role of boundary conditions in such problems and the complexity of their unambiguous choice \citep{warnecke2016influence}. At the same time, the role of asymmetry was emphasized in this type of problems both in the propagation of a magnetic field \citep{viviani2018transition} and in solving problems of the stability of the generated magnetic-field components stars \citep[see, e.g.,][]{bonanno2008non, bonanno2008stability}. These recent numerical experiments for solar-like stars have made it possible to better understand the relationship between poloidal and toroidal structures both in dynamo cycles and in the magnetic field of interstellar space \citep{bonanno2016stellar}.

It is obvious that today the most known and investigated star is the Sun, which demonstrates the characteristic periodic changes of the shape and direction of its large-scale magnetic field. Interplanetary scintillation observations from 1985 to 2013 year allowed investigate the North--South asymmetry of the global distribution of the solar-wind speed and the related magnetic field. \cite{2015JGRA..120.3283T} showed that the notable north-south asymmetry of polar solar winds occurs at solar maxima, and a small but significant asymmetry exists at solar minima. Also some evidence was found for the connection of the stellar cycle and magnetic topology \citep{2016MNRAS.462.4442S}. The peculiarity of the heliospheric global structure is the existence at low latitudes of the aforementioned heliospheric current sheet \citep{1961ApJ...134...20P, 2001A&A...376..288I, 2013SSRv..176..217R}, which is the extension of the streamer belt stretched along the magnetic neutral line in the solar corona \citep{1999SoPh..188..277E, 2001JGR...10615819S}. The dominance of a quadrupole magnetic field during solar maxima can be a reason of the appearance of two large-scale current sheets in the heliosphere, which was demonstrated recently by studies of scattered Lyman-$\alpha$ radiation of the Sun \citep{2020PhyU...63..801P} and was confirmed by model calculations \citep[e.g.][]{kislov2019quasi, maiewski2020magnetohydrodynamic}. Therefore the position and shape of the large-scale heliospheric current sheet in the whole heliosphere can carry the information about its origin in the interior of the Sun. Recently the role of large-scale current sheets as tracers of the internal Sun’s magnetic field and dynamo processes was proposed and discussed by \cite{ sym12122085}.

There is no doubt that the origin of solar current sheets is associated with cyclic solar magnetic activity, which in turn is believed to be driven by classical stellar dynamo. That dynamo acts somewhere in the solar interior, based on differential rotation and mirror asymmetry of physical processes in the stellar convective zone. However, particular features of the dynamo as well as magnetic-field distributions in stars remain a debatable accessory topic, demanding special investigations in each particular case.

\section{Statement of the Problem}
From the viewpoint of solar-dynamo studies as well as from the solar butterfly diagrams, it is known that the magnetic configuration in solar-type stars can be considered as a wave of quasi-stationary magnetic field propagating somewhere inside. This magnetic configuration contains both poloidal and toroidal components, wherein the latter hides in the convective shell and seems to be even stronger than the former. In contrast, from the solar current-sheet studies, the solar magnetic configuration can be considered as a combination of several temporally oscillating multipoles of dominantly poloidal magnetic field. The above difference of views on the one object, i.e. solar magnetic configuration, can be compared to the two viewpoints on radio-wave propagation. A radio wave looks quite different when being considered close to a source where the wave is excited and that in the wave zone. 

The problem of how to match the two viewpoints deserves clarification, which might be slightly of academic interest for the solar case. The point is that we get a lot of observational information separately for the solar magnetosphere (say, spacecraft data) and  cyclic magnetic activity in solar interior (say, sunspot data) and can postpone this matching for a future time. The scientific situation becomes quite different in the context of the contemporary break-through with extraterrestrial studies. We are now interested in understanding of magnetic features of magnetic configurations in various exoplanetary systems and have to face a drastic shortage of relevant observational information.  It looks reasonable to combine available knowledge, concerning a possible magnetic-field structure in the interior of the host star and that of the stellar current sheet. The aim of our study is to contribute in clarification of this problem.  

We depart from the statement that the solar analogy remains useful here; however we cannot fully base the study on this analogy. The point is that the dynamo action in a spherical body can excite various magnetic configurations apart from the solar one. In particular, the magnetic wave can in principle propagate from stellar equator to the stellar pole rather from the pole to the equator as it happens on the Sun. It looks reasonable to expect that the rich variety of stars known for contemporary astronomy can provide a possibility for dynamo to demonstrate various examples of dynamo-driven configurations known from dynamo modeling, and the main difficulty is how experimentally identify the configurations. In this context we would like to learn what difference in the current sheet behaviour we would expect for various magnetic configurations.

Another point to be clarified is that, as we learn from the famous Hale polarity low, magnetic field in solar interior is almost antisymmetric with respect to the solar equator (i.e. it has dipole-type symmetry in respect to the solar equator). According to the spacecraft observations as well as surface magnetic tracers, however, the magnetic configuration in the solar magnetosphere contains a pronounced quadrupole component, which has a non-dipole symmetry. It is natural to believe that this violation of dipole-type symmetry is associated with some moderate asymmetry of physical properties in respect to the stellar equator. A verification and quantification of this interpretation looks interesting in the stellar context: we are going to learn what degree of asymmetry in intensity of dynamo drivers is sufficient to get a stellar quadrupole moment that is realistic given our experience of solar observations.

In our article we tried to investigate the properties of dynamo processes in some stars available to produce magnetic fields similar to (or different from) the solar one and then extending through the whole astrosphere. For this aim we used two interrelated models describing different spatial scales: i) the simplified dynamo model and ii) the model of a transition layer where the magnetic field generated by dynamo mechanism floats into the higher layers of the stellar atmosphere and is transformed into the large-scale magnetic field of the star where the plasma flow is accelerated and becomes a stellar wind with known characteristics. Also we rely on the results of our previous model of the solar wind in the heliosphere \citep{maiewski2020magnetohydrodynamic, 2020PhyU...63..801P}, in which the setting of boundary conditions on some spherical surface around the Sun almost unambiguously determined the corresponding structure of the magnetic field and current sheets in the whole heliosphere. We assume that unique physical processes of a magnetic-field generation take place in the interiors of the Sun and many other stars. Therefore the use of such analogy allows us to investigate all chains of processes transmitted from small to large stellar scales and finally leading to the formation of a large-scale astrospheric current system. Formation of structures such as the heliospheric current sheet and other kinds of current sheets in space presumably depends on both the symmetry of dynamo processes inside stars and on the dynamics of large-scale magnetic fields in their environment. Naturally, in our work we will not be able to cover all of the diversity of the known classes of stars. Here we use the analogy with dynamo processes in the Sun as the most studied among other celestial objects and also consider possible dynamo mechanisms that may be realized in the Sun but can be considered for other stars.

\section{Dynamo Equations for Migrating Dynamo Wave}
We use the classical Parker's migratory dynamo -- one of the most simple MHD models, which describes main features of the dynamo process in thin spherical layers \citep{parker1955hydromagnetic}. Historically this model was developed to describe the magnetic field of the Sun; however, one would like to believe that it can correctly describe the general features of magnetic fields in stars with similar thin convective shells. Of course, it is possible to use more modern and detailed models, matched, for example, to the Sun's magnetic field; however, here we are going to consider the problem from the viewpoint of physical principles and consider the most simple physical model for stellar dynamo and the dynamo--magnetosphere matching. Our point is that it is reasonable to explore first the simple cases in the format of a physical article and only then move farther to more realistic models in format of an astronomical work.

The Parker migratory dynamo is based on the magnetic induction equation, averaged over a random velocity field. This averaging is usually carried out under the assumption of two-scale structure of the plasma turbulence; however, this requirement is optional. So further we assume that there are small-scale rapidly changing fluctuations with so-called nonzero hydrodynamical helicity $\langle\vec{v}\cdot{\rm curl}(\vec{v}) \rangle\ne 0$ and a large-scale, slowly varying field with a nonzero differential azimuthal rotation $\partial_r\Omega\ne 0$. Exactly these two features of the convective velocity make possible the realization of a magnetic dynamo and the formation of a migratory dynamo wave, described, e.g. by \cite{1980mfmd.book.....K}. In our investigation we use data obtained by Parker's model as a boundary condition for the problem of magnetic-field transfer in the exterior area. Varying North--South helicity distribution and signs of helicity and differential rotation, we restore symmetrical and asymmetrical migrating waves, propagated to the solar poles or equator. 

Note also that for a long time it was presumed that the mirror asymmetry of the flow (presented by hydrodynamical helicity) arises due to Coriolis force action, see, e.g. \cite{parker1955hydromagnetic}. Now people believe that the magnetic force is better to get the mirror asymmetry at least in the solar case (so-called Babcock--Leighton scheme) and other dynamo drivers like meridional circulation; however these important developments of solar-dynamo studies become important for more developed stages of modeling while here we stay with the simplest cases only.

After averaging over uniform and isotropic velocity field the magnetic induction equation can be written in the well-known Parker's form 
\begin{equation}
\partial_t \vec{B}=\rm{curl} \left(\alpha\vec{B}+ \vec{V}\times\vec{B}-\beta{\rm{curl}}(\vec{B})\right),
\label{Pareq}
\end{equation}
where $\vec{V}$ is an averaged velocity $\langle \vec{v}\rangle=[\Omega,\vec{r}]$, $\alpha$ is a hydrodynamical helicity, defined by the averaged scalar product: $-(\tau/3)\langle\rm{curl}(\vec{v})\cdot\vec{v}\rangle$ and $\beta$ is a turbulent diffusivity, connected with magnetic conductivity and hydrodynamical energy of the convective flow. Parker shows this equation in the azimuthally symmetrical spherical coordinate system is convenient to rewrite, decomposing magnetic field in a sum of poloidal and toroidal components:
\begin{equation}
\vec{B}=B\vec{e}^{\varphi}+r_0{\rm curl}(A \vec{e}^{\varphi}),
\end{equation}
where $r_0$ is the typical radius of the star's convective shell. In such a case the considering process in the thin shell defined only by two simple equations:
\begin{equation}\label{Eq03}
\partial_t A=R_{\alpha}\sin\theta\cos\theta B+\partial_{\theta}^2 A-\mu^2 A\, ,
\end{equation}
\begin{equation}\label{Eq04}
\partial_t B=R_{\omega}\,\partial_{\theta}(\sin\theta A)+\partial_{\theta}^2 B-\mu^2 B\, ,
\end{equation}
where time $t$ is measured in units of $r_0^2/\beta$, $\theta\in [0,\pi]$ and the radial part of diffusion is presented in simplified form, without $r$-derivatives, see, e.g. \cite{kleeorin2016dynamics}:
\begin{equation}
\frac{R^2}{r}\partial^2_r(Br)\sim -\mu^2 B.
\end{equation}
Parker's system (Equations 3\,--\,4) can be solved numerically, and the behaviour of its solution is defined by two 
dimensionless parameters:
\begin{equation}
R_{\alpha}=\frac{r_0\tau}{\beta} \langle\vec{v}\cdot{\rm curl}(\vec{v}) \rangle\quad\text{ and }\quad R_{\omega}=\frac{r_0^3}{\beta}\partial_r\Omega.
\end{equation}
These parameters are nonzero due to our assumptions about helical convection and differential rotation; moreover further we suggest that they are sufficiently large for generation, because the dynamo process is a threshold effect, below which only magnetic-field decay can be observed. So following the earlier works, e.g. \cite{kleeorin2016dynamics}, we get $\mu=3$ and $R_{\omega}=\pm 10^4$. Note that here we roughly suppose that $R_{\alpha}$ is North--South antisymmetrical, that provided by term $\cos\theta$ in (\ref{Eq03}), and maximal helicity is localized near the middle latitudes, that provided by $\sin\theta$. The term $\sin\theta$ has another advantage: considering North--South asymmetry we take $R_{\alpha}=1+\delta$ in the north hemisphere and $R_{\alpha}=1-\delta$ in the southern one, parameter $0<\delta<1$ is responsible for problem  asymmetry and $\sin\theta$ ensures continuity of helicical properties. Note also that dynamo problems are usually characterized by the $D$-parameter, where $D=R_{\alpha}R_{\omega}$, so for convenience we will describe further an asymmetry by the ratio
\begin{equation}
d=\frac{|D_{\rm north}-D_{\rm south}|}{D_{\rm north}+D_{\rm south}},
\end{equation}
where the subscript is responsible for the North--South hemisphere and designate cases with $R_{\omega}=10^4$ by $D^+$, and cases with $R_{\omega}=-10^4$ by $D^-$. That means, for example, that for the symmetrical case $D^+$ the parameter $d$ would be zero and $R_{\alpha}R_{\omega}=10^4$, while for the case $D^-$ with $d=-1\,\%$ parameter $R_{\omega}$ would be $-10^4$ and $R_{\alpha}=1,01$ ($0,99$) for the northern (southern) hemisphere. 

Calculating $A(\theta,t)$ and $B(\theta,t)$, we define magnetic field $\vec{B}(\theta,t)$ on the boundary stellar sphere, assuming weak $r$-dependency of these functions near the sphere boundary:
\begin{equation}
\vec{B}=\left(\frac{\partial_{\theta}(A\sin(\theta))}{\sin(\theta)},-r_0A,B\right)
\end{equation}
Finally, note that the linear mean-field system describing magnetic-field generation can have only exponentially growing solutions; to obtain a stabilized dynamo wave with saturation, we add a phenomenological nonlinear effect, usually called in MHD-dynamo theory $\alpha$-quenching. In other words, we assume that $R_{\alpha}$ decreases simultaneously with magnetic-field growth like
\begin{equation}
R_{\alpha}= \frac{1\pm\delta}{1+{\rm max}(|\vec{B}|)^2}.
\end{equation}
In this nonlinear suppression we take the maximum of magnetic field $[{\rm max}(|\vec{B}|)]$ over $\theta$ to remove redundant $R_{\alpha}$-dependency on the latitude, saving only North--South $\theta$-asymmetry. Examples of dynamo solutions, obtained after stabilization (for dynamo waves with not growing amplitudes), presented in Figure 1. On the left panel, the negative $D^-$ symmetrical case is demonstrated, which corresponds to equatorward dynamo wave, while the right panel corresponds to the symmetrical migrating wave with positive $D^+$, propagated to the poles.

\begin{figure} 
\centerline{\includegraphics[width=1\textwidth,clip=]{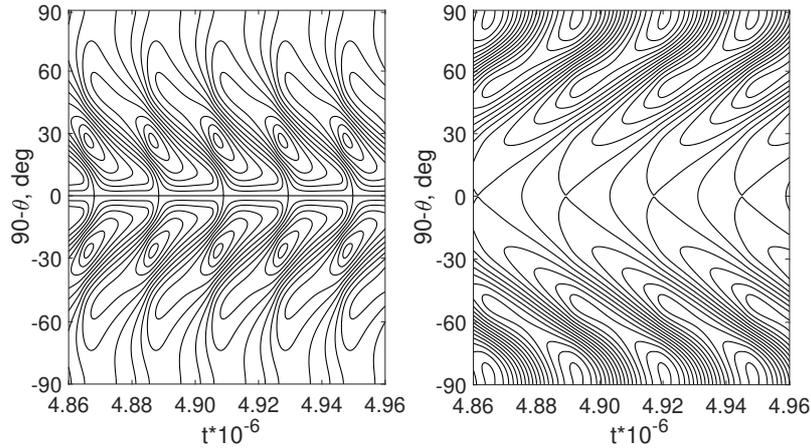}}
\caption{Examples of symmetrical $d=0$ migrating dynamo waves, obtained after stabilization. Butterfly diagrams show the level lines of the radial component of the magnetic field. On the {\it vertical axes} the latitude $90^\circ-\theta$ and on the {\it horizontal axes} the time $t$ are shown. The {\it left panel} is for the negative case $D^-$ and the {\it right panel} is for the positive case $D^+$.}\label{fig:epsart}
\end{figure}

Note that Equation ~\ref{Pareq} belongs to the transport equations, which usually describe diffusion or advection, while here we deal with magnetic-field self-excitation. The point is that here we deal with a vector quantity $\vec{B}$. The induction effect associated with differential rotation produces a poloidal magnetic field from a toroidal one, while mirror asymmetry restore toroidal magnetic field from poloidal one, which taken together, give magnetic-field self-excitation according to Parker's idea. Considering below magnetic-field propagation out of the dynamo active region we deal with magnetic-field propagation from the dynamo active region rather than with its self-excitation.

We stress again that the Parker migratory dynamo is an obvious oversimplification and ignores many important features of solar dynamo, such as meridional circulation among many others. There is no problem to include such effects and many much more realistic dynamo models are suggested in the contemporary scientific literature; however, here we deliberately stay with the simplest one just to demonstrate physical effects under discussion.

\section{Extension of the Magnetic Field From the Dynamo Region into the Transition Zone}
There exists an important difference between the magnetic field in the dynamo-active region inside the Sun and the one propagating in the heliosphere. Magnetic diffusion is weak in the dynamo-active region. Magnetized matter exists in the dynamo-active region for quite a long time. This time can be sufficient for the formation of two flux ropes with oppositely directed magnetic fields. Ohmic losses for such ropes are small but not negligible, but as a result of magnetic-field propagation outwards, dissipation processes can lead to almost immediate annihilation of oppositely directed magnetic flows with the fast destruction of the total magnetic flux. Taking this into account, we have constructed the model of a transition zone between the internal stellar dynamo zone and the external boundary (usually it is the Alfv\'<en surface or a more distant closed surface) where the magnetic field of convective shells can diffuse and annihilate; as a result large-scale magnetic field can be formed near the external boundary, which further propagates in an almost unchanged form over the whole astrosphere. We supposed that the transition zone has the shape of a spherical layer bounded by spheres with radii $r_0<r<r_1$, where $r_0$ is the radius of the internal boundary of the transition zone (the boundary of convective shells) and $r_1$ is the radius of the outer sphere.

Examples of such effects can be found in a series of exactly solvable dynamo models at the early stages of dynamo studies \citep[see, e.g.,][]{zeldovich1983magnetic} to demonstrate that diffusion terms are inevitable in the dynamo modeling. In contrast, magnetic field propagates from the dynamo active region into the heliosphere in a rather short time, and effects of catastrophic cancellation are unknown in this context. Therefore we suppose the following properties of magnetic-field solutions in the transition zone: 

\begin{enumerate}[i)]
    \item The magnetic field $\vec{B}$ on $r=r_0$ in a steady state at long times is a $T$-periodic function of time $t$.
    \item At the internal boundary sphere $r=r_0$  only the radial magnetic field $B_r$ can be taken into account because of the smallness of other magnetic components $B_\theta, B_\varphi$ in comparison with the radial one.  
    \item In the transition zone $r_0< r < r_1$ the magnetic field $\vec{B}$ can be described by the magnetic induction equation without convective term:
\begin{equation}\label{Eq10} 
\partial_t \vec{B} = -\eta\, {\rm{curl}} ({\rm{curl}}\,\vec{B})\, ,
\end{equation}
where the coefficient $\eta$ is responsible for convective diffusion. 
    \item The propagation of the magnetic field outside the transition zone, i.e. at $r>r_1$, is assumed to satisfy the magnetic induction equation without  the diffusion term:
\begin{equation}\label{Eq12}
\partial_t \vec{B} = \rm{curl} (\vec{V}\times \vec{B}).
\end{equation}
This is a typical modeling assumption for the description of the solar wind in the heliosphere (and perhaps of stellar winds of solar-like stars in astrospheres). We should note that when we consider the equation of the magnetic-field propagation outside the transition area in the heliosphere the diffusion term becomes negligibly small \citep{kislov2019quasi, maiewski2020magnetohydrodynamic}. 

To reconcile the third and fourth assumptions, the magnetic field on the outer boundary sphere $r=r_1$ should satisfy the equation
\begin{equation}\label{Eq13}
\partial_t B_r = \frac{1}{r\sin\theta} \partial_{\theta}\Bigl(\sin\theta \bigl(V_r B_{\theta} - V_{\theta} B_r\bigr)\Bigr) \,.
\end{equation}

    \item We assume that the velocity distribution on the outer sphere $r=r_1$ satisfies the conditions
\begin{equation}
V_{\theta} = 0, \quad V_r =\mathrm{const}.
\end{equation}
These conditions can be taken as the simplest model of stellar-wind propagation near the outer sphere.

\end{enumerate}

Let us note that the distribution of the radial magnetic field $B_r$ at the inner boundary $r=r_0$ of transition zone, obtained from the dynamo model (described in paragraph 3), 
can be decomposed into a double Fourier--Legendre series on time $t\in[0; T]$ and angle $\theta\in(0; \pi)$ as follows:
\begin{equation}\label{Eq11}
B_r\Bigr|_{r=r_0} = \sum\limits_{n=1}^{\infty} \sum\limits_{k=-\infty}^{\infty} \gamma_{nk} l_n(\cos\theta) \mathrm{e}^{\mathrm{i}\omega kt},
\end{equation}
where $l_n(z)$ are Legendre polynomials: $l_1=z$, $l_2=(3z^2-1)/2$, \ldots; $\mathrm{i}$ is an imaginary unit, $\gamma_{nk}$ are coefficients and $\omega=2\pi/T$, where $T$ is the period of $\vec{B}$. 
	
To solve Equation \ref{Eq10} with boundary conditions given by Equations \ref{Eq13} and \ref{Eq11}, we decomposed $\vec{B}$ into a Fourier series in time $t$ in $[0, T]$:
$$
\vec{B} = \sum\limits_{k=-\infty}^{\infty} \vec{B}^k \mathrm{e}^{\mathrm{i}\omega kt}.
$$
Substituting this decomposition in Equation \ref{Eq10} one can obtain a system of equations for the coefficients
\begin{equation}\label{Eq14}
\frac{\mathrm{i}\omega k}{\eta}\, \vec{B}^k = - \rm{curl} (\rm{curl}\, \vec{B}^k)\quad\text{ and }\quad
\div \vec{B}^k = 0,
\end{equation}
which can be integrated using potentials $u_k$ in the form
$$
B^k_r = -\frac{1}{r^2\sin\theta}\, \partial_{\theta} u_k,
\quad\text{ and }\quad
B^k_{\theta} = \frac{1}{r\sin\theta}\, \partial_r u_k.
$$
Then the first equality of Equation \ref{Eq14} is transformed into the equation for $u_k$
\begin{equation}\label{Eq15}
\frac{\mathrm{i}\omega k}{\eta}\, u_k = \partial_r^2 u_k + \frac{\sin\theta}{r^2}\,\partial_{\theta}\Bigl(\frac{1}{\sin\theta}\,\partial_{\theta} u_k\Bigr)
\end{equation}
and the equation for the component $B^k_{\varphi}$. 

Then we solve Equation \ref{Eq15} by separation of variables
$$
u_k(r,\theta) = \sin^2\theta\, \sum\limits_{n=1}^{\infty} \Bigl(\alpha_{nk} X_{nk}^{+}(r) + \beta_{nk} X_{nk}^{-}(r)\Bigr) l'_n(\cos\theta),
$$
where $X_{nk}^{+}$, $X_{nk}^{-}$ are two linearly independent solutions of the equation
\begin{equation}\label{EqX}
r^2 X'' - \Bigl(n(n+1) + \frac{\mathrm{i}\omega k}{\eta} \,r^2\Bigr) X = 0,
\end{equation}
having the following asymptotics at large $\eta$ and limited $k,r$
\begin{gather*}
X_{nk}^{+}(r) = r^{n+1} \Bigl(1 + \frac{\mathrm{i}\omega k}{\eta} \, \frac{1}{2(2n+3)}\, r^2 + \ldots\Bigr), \\
X_{nk}^{-}(r) = r^{-n} \Bigl(1 - \frac{\mathrm{i}\omega k}{\eta} \, \frac{1}{2(2n-1)}\, r^2 + \ldots \Bigr). 
\end{gather*}
Finally we find the radial and meridional magnetic field in the transition zone:
\begin{multline}\label{Br_res} 
B_r(t,r,\theta) = -\frac{1}{r^2}\, \sum\limits_{n=1}^{\infty} \sum\limits_{k=-\infty}^{\infty} 
n(n+1) l_n(\cos\theta)\,  \mathrm{e}^{\mathrm{i}\omega kt}
\Bigl(\alpha_{nk} X_{nk}^{+}(r) + \beta_{nk} X_{nk}^{-}(r)\Bigr),  
\end{multline}
\begin{multline}
B_{\theta}(t,r,\theta) = \frac{\sin\theta}{r}\, \sum\limits_{n=1}^{\infty} \sum\limits_{k=-\infty}^{\infty} l'_n(\cos\theta)\,  \mathrm{e}^{\mathrm{i}\omega kt}
\Bigl(\alpha_{nk} {X_{nk}^{+}}'(r) +\beta_{nk} {X_{nk}^{-}}'(r)\Bigr),
\end{multline}
where $\alpha_{nk}$ and  $\beta_{nk}$ are coefficients of decomposition.

The boundary condition (Equation \ref{Eq13}) can be transformed to the form
\begin{equation}\label{Eq20}
\partial_r u_k + \frac{\mathrm{i}\omega k}{V_r} \, u_k = 0.
\end{equation}
As a result, the conditions in Equations \ref{Eq11} and \ref{Eq20} give the system of equation for coefficients $\alpha_{nk}$ and  $\beta_{nk}$. 

But what happens in the outer region, i.e. astrosphere? Here we can use the analogy with the known characteristics of the heliosphere. After finding both the solution inside the transition zone and the boundary conditions on its outer boundary we can make a prediction about the shape of neutral surfaces (or large-scale current sheets) in the whole heliosphere or astrosphere. Such numerical investigations were made in earlier models \citep[e.g.][]{kislov2019quasi, maiewski2020magnetohydrodynamic} of the solar wind in heliosphere. Here the sphere enclosing the Alfven surface was considered as the boundary surface where the magnetic components were set for common-sense reasons. It was shown that for large $r$ the value of $B_r r^2$ practically does not depend on $r$ \citep[][]{kislov2019quasi, maiewski2020magnetohydrodynamic}. On the base of earlier modeling, after crosslinking the boundary conditions in the inner (dynamo), transition, and outer region (astrosphere), we assumed that the condition $B_r r^2$ is valid in the region $r\ge r_1$. Consider the surfaces of the zero radial field $B_r=0$. Due to the axial symmetry of the model and the assumption $B_r\sim r^{-2}$, these surfaces have almost conical shape. Because the neutral surfaces in the solar wind  $B_r=0$ correspond to large-scale current sheets \citep[][]{1982SoPh...77..363L, 1991A&A...243..492V}, we can conclude that the neutral surfaces have the following properties: they should have disk-like shapes at low latitudes and conical shapes at the high latitudes. Moreover, as it was shown by \cite{ maiewski2020magnetohydrodynamic}, the shapes of neutral surfaces (and corresponding current sheets) can evolve in time, moving from low latitudes to higher ones and otherwise, accordingly to the temporal behavior of a helio-magnetic field during the solar activity cycle.

\section{Selection of the Main Parameters} 
Numerical values of the period $T$ and coefficients $\gamma_{nk}$ of the expansion given in Equation \ref{Eq11} for $n=1,\ldots,10$ and $k=-10,\ldots,10$ were obtained from the steady-state numerical periodic solution of the dynamo model. Figure 2 shows the characteristic periods $T$ of dynamo waves as a function of the asymmetry coefficient $d$  for dynamo models with the negative $D^-$ (left panel) and positive $D^+$ (right panel) dynamo numbers. One can see that the period of dynamo waves for negative $D^-$ decreases with the increase of the asymmetry coefficient; on the contrary for positive $D^+$ it increases. We see that generally the period $T$ depends weakly on the asymmetry coefficient $d$).

\begin{figure} 
\centerline{\includegraphics[width=1\textwidth,clip=]{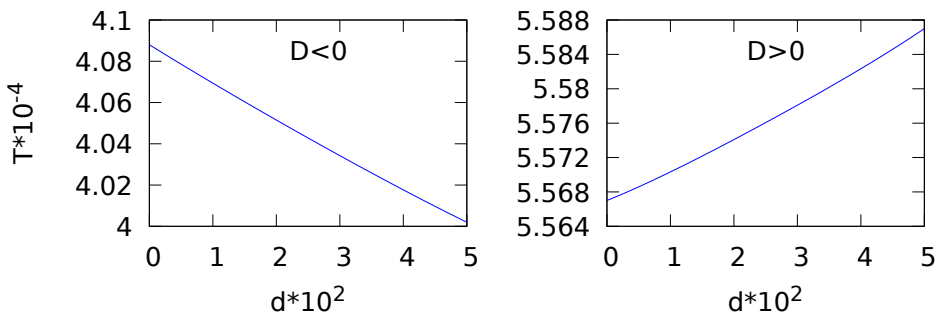}}
\caption{Period $T$ for various dynamo models. On the {\it horizontal axes} the asymmetry coefficient $d$ and on the {\it vertical axes} the dimensionless periods $T$ (in the units of $r_0^2/\beta$) are shown. The {\it left panel} is for the negative case $D^-$ and the {\it right panel} is for the positive case $D^+$.}\label{fig:period}
\end{figure}

Our calculations showed that the coefficients $\gamma_{nk}$ decrease exponentially with fixed $n$ and increasing $|k|$. Note that the even harmonics (even $k$) are practically absent, and for odd $ k $ the coefficient $\gamma_{nk}$ decreases approximately as $0.5^{|k|}$. The absence of even harmonics means that the radial magnetic field changes sign every half period. Therefore, the first harmonic $|k|=1$ is the main one and the influence of the other harmonics can be neglected to a rough approximation.

Here and below we will consider the estimates of the characteristics for Sun-like stars. If we assume that $r=1$ corresponds to the solar radius of $7\times 10^5$ km, and the period $T=4\times 10^4$ (for the dynamo model with negative $D^-$) corresponds to the $22$-year period of the solar activity, then $V_r=10$ corresponds to the average solar wind speed about $400$ ${\rm kmS^{-1}}$. The value $\eta$ is responsible for convective diffusion. For solar granules and supergranules it is of the order $10^9$ ${\rm{m^2S^{-1}}}$. We used the value $\eta=0.1$ in our units.

\section{Results}
We perform the above presented calculations in order to obtain magnetic configuration in the dynamo-active region and in the stellar neighborhood. Obviously, the configurations depend on various governing parameters. Our aim is to isolate stable and instructive features in the bulk of the obtained configurations. We avoid presenting small variations of the configuration related to parameter variations and focus attention on the features that look instructive. First of all, the configurations with positive and negative dynamo numbers $D$ behave quite differently. As expected, for $D^-$ the dynamo propagates equatorwards while for $D^+$ the wave propagates polewards. It is however far from the only difference between two cases.

For the negative $D^-$ toroidal and radial magnetic fields are antisymmetric relative to the stellar equator, which corresponds to the solar case. Such symmetry is known in dynamo studies as the dipole; however, the magnetic field as a whole is far from being just a magnetic field of magnetic dipole.  Odd magnetic multipoles have dipole symmetry in respect to the stellar equator. In fact, octupole provides a visible contribution in all solutions obtained. In contrast, for the positive $D^+$ the toroidal and radial magnetic fields are symmetric with respect to the solar equator. The configuration is known as the quadrupole one. The option possibility that dynamo-driven magnetic field may be of dipole and quadrupole symmetry is well known in dynamo modeling and discussed in the literature \citep[see for review][]{Metal08}. Of course, higher even multipoles contribute in magnetic fields of quadrupole type.  

Whether we obtain quadrupole or dipole configuration in a particular dynamo model depends on fine tuning of the governing parameters \citep{Metal08} and dipole configuration is obtained for the equatorward propagating wave because we depart from the solar phenomenology. It is not a problem to play here with numbers however we avoid giving too many plots here.

Quite unexpectedly, the quadrupole configuration is much more robust than the dipole one. We reduce the number of plots for the configuration to save space and concentrate attention on instructive features. Of course, quadrupole configurations depend on the asymmetry $d$; however, this dependence is weaker rather for the dipole one.

\subsection{Neutral Surfaces of the Radial Magnetic Field and Corresponding Current Sheets} 
\begin{figure} 
\centerline{\includegraphics[width=1\textwidth,clip=]{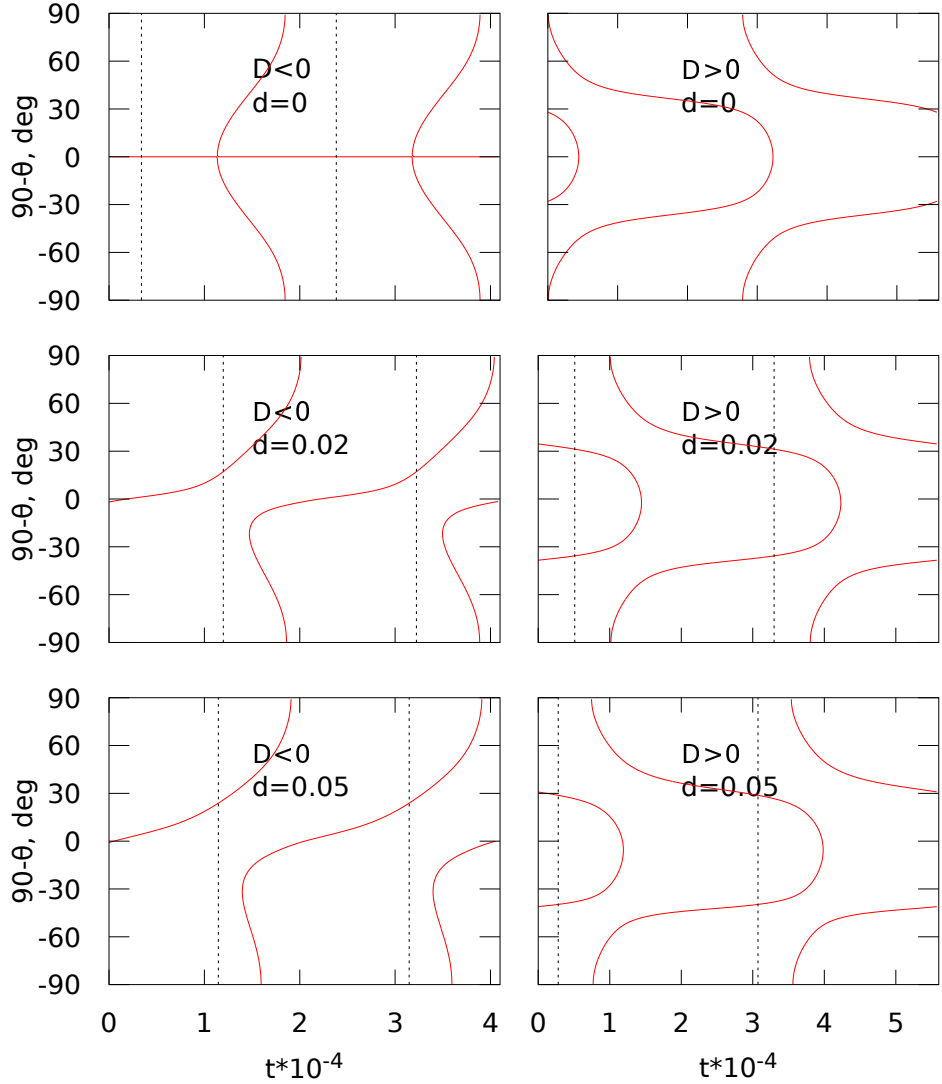}}
\caption{The location of the neutral surfaces $B_r=0$ for various dynamo models. {\it Left column} corresponds to the dipole configuration [$D^-$] and {\it right column} corresponds to the quadrupole one [$D^+$]. {\it Upper row} presents configurations with the exact hemispheric symmetry. {\it Middle and lower rows}  present  configurations with the asymmetries $d=0.02$ and $d=0.05$. The dashed lines indicate moments when the absolute value of amplitude multiplier of dipole component of $B_\varphi$ on the sphere $r=1$ reaches the maximum. On the {\it horizontal axes} the dimensionless time $t\times 10^{-4}$ and on {\it vertical axes} the latitude $90^\circ-\theta$ are shown.}\label{cursh}
\end{figure}

First of all, we consider location of current sheets in comparison with values of the dynamo drivers. We recall that the location of current sheets in the stellar neighbourhood is determined by the neutral surface of the radial magnetic field (Figure~\ref{cursh}). The location of the neutral surfaces depends on the chosen dynamo model, i.e. on the sign of $D$ and $d$, the radius of the outer sphere $r_1$ and the phase of the stellar cycle. For illustration, the value $r_1=2.85$ was chosen, for which the model corresponds to the Sun (see the following subsections).

Figure~\ref{cursh} shows the positions of the neutral surfaces $B_r=0$ depending on the time when dynamo waves propagate in the magneto-active zone and the evolution of the stellar magnetic field takes place. The results of the dynamo model with negative $D^-$ are shown in the left column of Figure~\ref{cursh} and the results with positive $D^+$ are presented in the right column of Figure~\ref{cursh}. The left column corresponds to the dipole magnetic configuration mixed with an octupole and the right column corresponds to the quadrupole magnetic component. The figures in the first line of the panel are characterized by the complete symmetry of dynamo waves in the northern and southern hemispheres ($d=0$). The magnetic configurations in the figures in the middle and lower rows were calculated at the asymmetry coefficients, respectively $d=0.02$ and $d=0.05$.

The mixture of dipole and octupole components of the magnetic field in different periods of the stellar cycle is characterized by the presence of one to three neutral surfaces. Therefore, in the initial moment of time for the symmetric case ($D^-,d=0$), the single neutral surface is located in the equatorial plane at zero latitude. At the time moment $t=1.2$ the equatorial neutral surface is bifurcated into three surfaces. Then two external neutral surfaces propagate to high latitudes, and the central one remains at zero latitude. In configurations with non-zero asymmetry coefficients $d=0.02$ and $0.05$, the configuration of neutral surfaces and their evolution is different. Thus at the initial moment of time the neutral surface is located at the zero latitude. Then it moves to higher latitudes in the northern hemisphere. At time $t=1.5$, the new neutral surface appears at the latitude of about $-20$ degrees, which, in turn, is split into two, i.e. the upper and the lower one. The upper surface (it occupies the middle position between three surfaces) begins to move from the southern hemisphere to the northern one, i.e. in the low latitude region, the second one descends to the south pole. At the moment when the median neutral surface reaches zero latitude, and simultaneously the northern and southern branches reach the poles, the topology of the neutral lines corresponds to the moment of time $t=0$, after which the magnetic evolution repeats.

As expected, we obtain equatorward propagating dynamo waves for the case representing the solar conditions. Location of this case in the parametric space of dynamo drivers agrees with that ones in solar dynamo modeling. In this sense our model agrees with standard ideas of the solar dynamo. We note here that quite unexpectedly we obtain that magnetic structures in the outer layer of the problem under consideration, i.e. current sheets, propagate polewards while magnetic structure in the dynamo active layer, i.e. the dynamo wave,  propagates equatorwards. This may be considered as a counterintuitive phenomenon and we did not find exactly that mentioned in the current scientific literature. In a broad sense, however, propagation of some details of magnetic configuration in direction opposite to the propagation of the main dynamo waves is known in dynamo studies.  In particular, \cite{KS95} found that dynamo waves in the simplest Parker model with a suitable profile of dynamo governing parameter propagates polewards in the polar vicinity while it propagates equatorwards in the main bulk of the dynamo-active shell.  

The different dynamics of neutral surfaces take place in the case of the dominating quadrupole field of the star (Figure 3, right column). At the initial moment of time, in the symmetric case ($D^+,d=0$) we see four neutral surfaces, two of which are located at latitudes $\pm 30$ degrees in the northern and southern hemispheres, and other two surfaces are situated at the poles of the star. Then these neutral surfaces move towards the Equator, where they connect in pairs at different points in time and then disappear. Thus, for different time periods of the magnetic cycle the presence of two or four neutral surfaces is characteristic, and this configuration is also cyclically repeated. In the case of asymmetry ($D^+, d=0.02, 0.05$), we should note that a phase shifted is seen in comparison with the symmetric case.

Generally, comparing the examples of neutral surfaces evolution in stellar astrosphere we conclude that each component of the multipole decomposition of the star's magnetic field contributes to the formation of the even or odd numbers of neutral surfaces (i.e. corresponding current sheets); these numbers depend on the parity of the harmonics itself. In the case of the mixed contribution of several harmonics to the stellar magnetic field, the number of neutral surfaces corresponds to the harmonics with the dominant contribution to the total magnetic field. 

\begin{figure} 
\centerline{\includegraphics[width=1\textwidth,clip=]{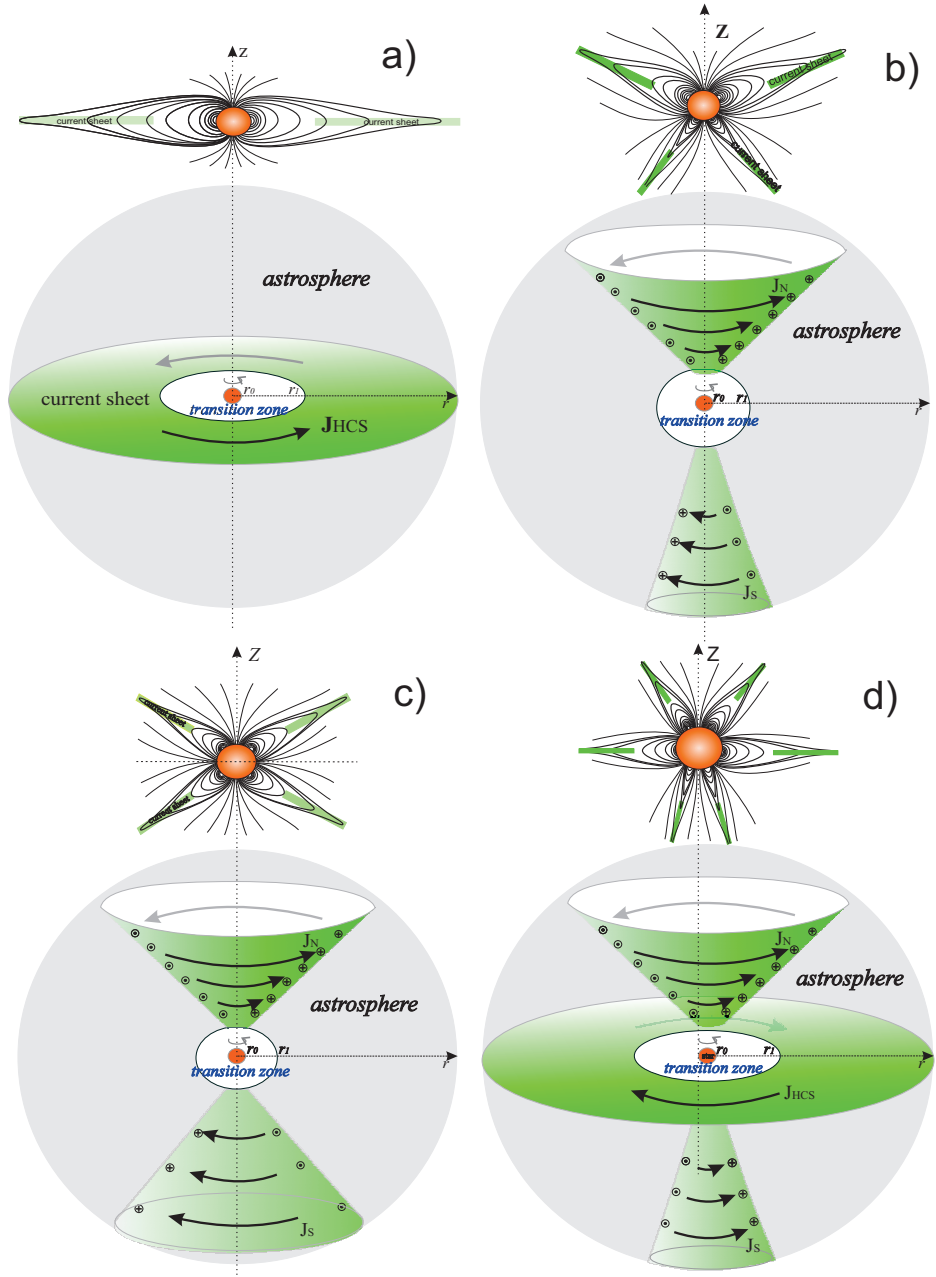}}
\caption{The schematic shapes of stellar magnetospheres with (\textbf{a}) dipole, (\textbf{b}) mixed dipole and quadrupole, (\textbf{c}) quadrupole and (\textbf{d}) mixed octupole and dipole/quadrupole magnetic fields that should determine the corresponding structure of large-scale current sheets in the astrospheres of Sun-like stars.
}\label{magnetosph}
\end{figure}

We mentioned above that large-scale current sheets in the heliosphere are the extension of neutral lines of the Sun. Thus, one should expect that this property has a unique character for Sun-like stars, and consequently their current sheets in astrospheres are located along the neutral surfaces related with the neutral lines in coronas. Figure~\ref{magnetosph} demonstrates some possible shapes and quantities  of large-scale current sheets in stellar astrospheres depending from the domination of corresponding magnetic components: dipole (Figure~\ref{magnetosph}a), mixed dipole and quadrupole (Figure~\ref{magnetosph}b), quadrupole (Figure~\ref{magnetosph}c) and asymmetric configuration of octupole mixed with quadrupole/dipole magnetic fields (Figure~\ref{magnetosph}d).  

At the top of several schematic astrospheres  the 2D view of corresponding stellar magnetospheres are shown. The multipole decomposition of the magnetic field contributes to the formation of even or odd numbers of neutral surfaces and corresponding current sheets. This depends on the parity of the harmonics themselves. In the case of the mixed contribution of several harmonics to the stellar magnetic field one can see two general effects: i) North-South asymmetry of current sheets distribution; ii) the number of neutral surfaces corresponds to the harmonics with the dominant contribution to the total magnetic field. 

Lets us consider Figure~\ref{magnetosph} as an example of the helio-magnetic field and the corresponding current-sheet configuration, accordingly  \citep{maiewski2020magnetohydrodynamic, 2020PhyU...63..801P}. Figure~\ref{magnetosph} shows the dipole magnetosphere of the quiet Sun with the equatorial neutral line as the source of a disk-like heliospheric current sheet ($J_{\text{HCS}}$). During the solar-activity cycle the dipole magnetic component decreases, while the quadrupole one increases. As a result the helio-magnetic field becomes asymmetric; the heliospheric current sheet moves to higher latitudes of the northern hemisphere and takes the shape of a cone. The greater the latitude, the narrower the conical sheet is. At the same time, at the high latitudes of the opposite hemisphere the new conical current sheet forms and begins to move to lower latitudes. The directions of such eddy currents in the northern and southern hemispheres are different; they are shown by arrows and small circles. The configuration shown in Figure~\ref{magnetosph}b is different from the one shown in Figure~\ref{cursh} at the bottom left where the third neutral line (influence of octupole component) goes down to the southern hemisphere. At the solar maximum period the quadrupole field is generally dominant, and as a result the magnetic field seems to be almost symmetrical, as is seen in Figure~\ref{magnetosph}c. Further more heliospheric current sheet goes towards the northern pole and ends its existence there, while the second current sheet occupies the equatorial region. As a consequence the directions of magnetic fluxes in the northern and southern hemispheres are reversed. 
	
The choice of current direction depends on the star’s dynamo characteristics; however, one rule must be true in cases of many sheets: currents of adjacent sheets should have opposite directions and magnetic fluxes on both sides of the neutral surfaces should be opposite. An asymmetrical current configuration in the case of the dominance of octupole magnetosphere mixed with quadrupole/dipole is shown in Figure~\ref{magnetosph}d, where one can see the heliospheric current sheet at low latitudes and two large-scale conical current sheets at the higher latitudes. This figure is in accordance with (Figure~\ref{cursh}, left) and simulations of stellar magnetic topology taking into account different multipole fields \citep{2015ApJ...798..116R}. 

In the previous paragraph we tried to study the fundamental problem of how the generation of multipole magnetic harmonics in the interiors of stars can affect the structure of their astrospheres. Returning to a variety of possible stellar magnetic configurations due to dynamo processes in stellar interiors we would note that the symmetry of stellar magnetic fields can be formed when the field has multipoles of only odd or only even orders. The mix of multipoles of different orders leads to the formation of asymmetric magnetic configurations in astrospheres. The dynamics of neutral surfaces with predominant quadrupole-octupole (Figure~\ref{cursh}, left) and purely quadrupole components (Figure~\ref{cursh}, right) illustrated in  Figure~\ref{magnetosph} can differ from the above-mentioned evolution of solar neutral surfaces \citep{maiewski2020magnetohydrodynamic, 2020PhyU...63..801P} for different stars. Particularly, the merging of two close neutral surfaces is possible, as shown in (Figure~\ref{cursh}, right). Such processes are not characteristic for the Sun's environment, but can be realized for other stars.

\subsection{Hemispheric Asymmetry}
As we mentioned above, the magnetic configurations for dynamo drivers that are exactly symmetric in respect to the stellar equator can be summed up as being of dipole or quadrupole type. As for the solar magnetic field, it obviously contains even magnetic multipoles in addition to the odd ones, i.e. strictly speaking it is a mixed-parity configuration. Admixture of even harmonics are visible in various tracers of magnetic activity; just as an example, it was investigated for zonal harmonics of surface solar magnetic field (see, e.g., the recent article by \cite{Oetal21}  and references therein). Here, however, we are interested in hemispheric asymmetry as recorded in heliospheric  data. To be specific we use here the data obtained by the {\it Ulysses} mission.

\subsubsection{Asymmetry of Solar Magnetic Field According to Ulysses Data}
Below we use {\it Ulysses} data to estimate the multipole harmonics of the helio-magnetic field in the different periods of the solar activity. Low and high-latitude regions of the heliosphere were explored by {\it Ulysses} (a joint project of the ESA and NASA, launched in 1990 and terminated in 2009), whose mission was to orbit the Sun and to study the physical characteristics of the solar environment at all latitudes, including the polar regions \citep{1991SmithE&S.....4...10S, 1992MonsignoriNCimC..15..493M}. Until now {\it Ulysses} remained the unique spacecraft that moved along a heliocentric orbit almost perpendicular to the ecliptic plane. {\it Ulysses} circled the Sun three times over its northern and southern poles. Its heliocentric orbit had perihelion at about 200 million km and aphelion at about ~810 million km, with a period of 6.2 years. During its mission {\it Ulysses} observed two solar cycle minima in 1996 and 2009, and solar maxima in 1990 and 2000. This spacecraft obtained valuable data on the structure and dynamics of the magnetic fields of the Sun \citep[e.g.][]{1995SmithGeoRL..22.3325S, 1998NeugebauerJGR...10314587N, 2012ManoharanApJ...751..128M, 2017KhabarovaApJ...836..108K}. The most important result of the {\it Ulysses} mission was the confirmation of the idea of the four-dimensionality of the heliospheric structure and dynamics, which depend not only on the spatial coordinates (as the distance from the Sun, helio-latitude and helio-longitude), but also on time  \citep{2013BaloghSSRv..176..177B, 2000DmitrievAdSpR..25.1965D}. 

Let us suppose that at large values of $r$ the radial magnetic field can be described by the dependence
\begin{equation}\label{Br_Uly}
B_r(t,r,\theta) = \frac{1}{r^2}\,\sum\limits_{n=1}^3 \delta_n l_n(\cos\theta)\cos(\omega t+\psi_n),
\end{equation}
where only the first harmonic $|k|=1$ is taken into account, and  $\delta_1, \delta_2, \delta_3$ are the amplitude multipliers of, correspondingly, dipole, quadrupole and octupole components, $\psi_1, \psi_2, \psi_3$ are phase shifts. Our approach is the development of an earlier attempt \citep{2012VeselSoSyR..46..149V}, made in the framework of a potential model, to describe the magnetic field of the Sun's corona and inner helio-magnetic field in the form of a sum of only dipole and quadrupole fields on the base of {\it Ulysses} data.  

Then, for the Sun, the values $\delta_n/\delta_1$, $\psi_n-\psi_1$ can be found from a comparison with the results of {\it Ulysses}/VHM\_FGM, see \href{http://ufa.esac.esa.int/ufa/#data}{http://ufa.esac.esa.int/ufa/data}:
\begin{gather}\label{Asun}  
\delta_2/\delta_1=0.20, \quad 
\delta_3/\delta_1=0.44, \\
\nonumber
\psi_2-\psi_1=-0.43\pi, \quad
\psi_3-\psi_1=0.23\pi.
\end{gather}
To obtain these data, we substituted the dependencies $r(t)$ and $\theta(t)$ corresponding to the orbit of {\it Ulysses} into the Equation \ref{Br_Uly} and, comparing them with the measurements of the spacecraft, calculated the coefficients by the least squares method. The evaluation of the coefficients (Equation \ref{Asun}) is presented for the first time according to the {\it Ulysses} data. We stress again that here we remain with the simplest approach to the topic under discussion and avoid possible development of the dependence (Equation \ref{Br_Uly}).

\subsubsection{Comparing Various Dynamo Models with Ulysses Data} 
At this stage we compare our theoretical conclusions with observations. The most reasonable idea seems to be a comparison of the dynamo parameters from the model with the observational data of the solar wind obtained by the {\it Ulysses}. We compare below the experimental estimates (Equation \ref{Asun}) with corresponding theoretical estimates obtained from dynamo modeling. The amplitudes $\delta_n$ and phase shifts $\psi_n$ for this comparison can be taken from the Equation \ref{Br_res} under the assumption that $B_r r^2$ is independent of $r > r_1$. 
Thus, we have
\begin{gather}
\label{deltan}
\delta_n = -2n(n+1)\Bigl|\alpha_{n,1} X_{n,1}^{+}(r_1) + \beta_{n,1} X_{n,1}^{-}(r_1)\Bigr|, \\
\nonumber
\psi_n = \arg\Bigl(\alpha_{n,1} X_{n,1}^{+}(r_1) + \beta_{n,1} X_{n,1}^{-}(r_1)\Bigr),
\end{gather}
where $X_{nk}^{\pm}(r)$ are defined by Equation \ref{EqX}) coefficients $\alpha_{nk}, \beta_{nk}$ depend on $r_1$ through Equations \ref{Eq11} and \ref{Eq20}. 

\begin{figure} 
\centerline{\includegraphics[width=1\textwidth,clip=]{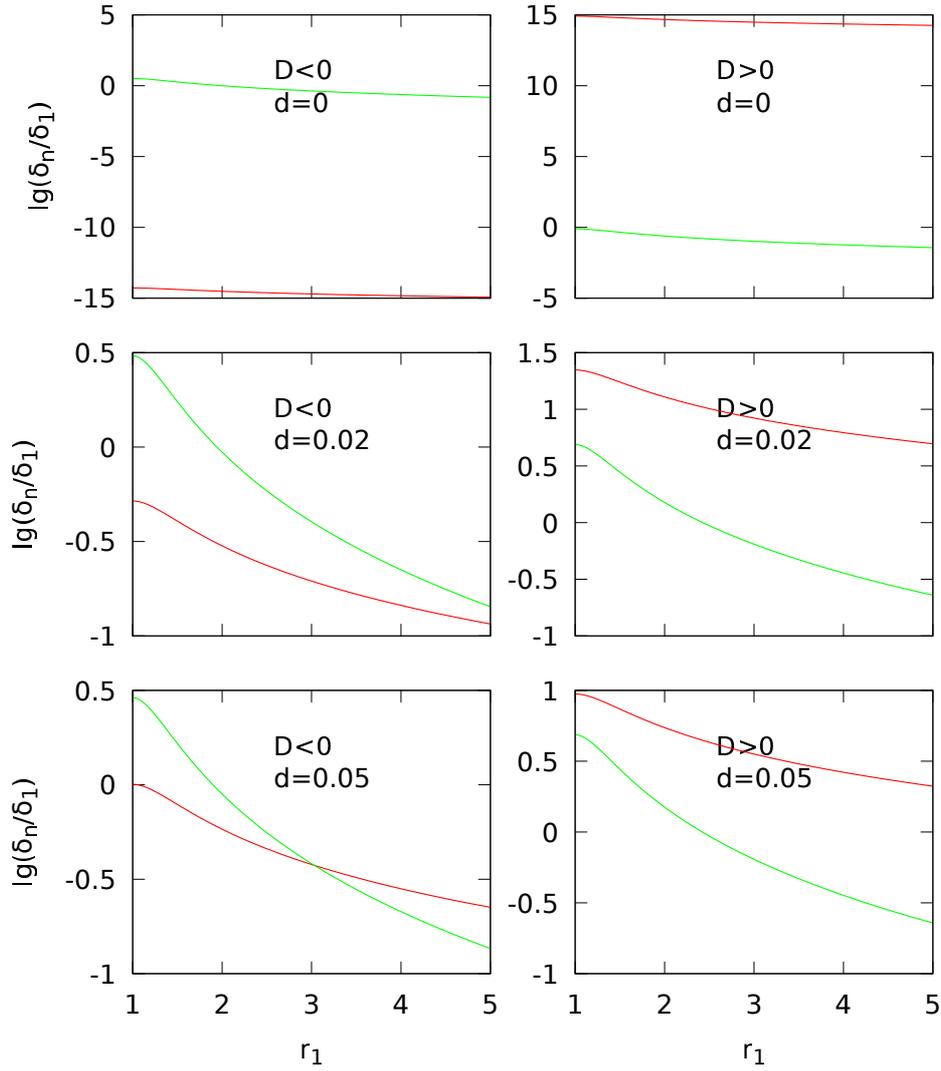}}
\caption{Dependencies $\log_{10}(\delta_2/\delta_1)$ (red) and $\log_{10}(\delta_3/\delta_1)$ (green) on $r_1$ for various dynamo models. {\it Left column} corresponds to a dipole configuration ($D^-$) and {\it right column} --- to a quadrupole one ($D^+$). {\it Upper row} presents configurations with exact hemispheric symmetry; {\it middle and lower rows}  present  configurations with asymmetries $d=0.02$ and $d=0.05$.}\label{fig:coeff_quot}
\end{figure}

Parameter $\delta_n/\delta_1$ from our dynamo model decreases with the growth of $r_1$ approximately as $r_1^{-(n-1)}$ as it follows from Equation \ref{deltan}, as is shown in Figure~\ref{fig:coeff_quot}. This is true for $1 < r_1 \ll \sqrt{\eta/(\omega k)}$. For $D^-$ and small $r_1$, the octupole prevails over the quadrupole, but the situation changes to the opposite with the growth of $r_1$. For $D^+$ the quadrupole prevails over the octupole for all considered values of $r_1$. The values of $\delta_3/\delta_1$ practically do not depend on $d$.

\begin{table}
\caption{The phase shifts $\psi_n-\psi_1$ in units of $\pi$ for various asymmetries. {\it In the first two columns} $D^-$, {\it in the last columns} $D^+$. In the case with $D^-$ and $d=0$ there is practically no quadrupole component, so $\psi_2-\psi_1$ is undefined. Similarly, in the case with $D^+$ and $d=0$ there are practically no odd multipoles.}\label{tab:phas_shft} 
\begin{tabular}{c c c c c }
\multirow{2}{*}{d} & $\psi_2-\psi_1$ & $\psi_3-\psi_1$ & $\psi_2-\psi_1$ & $\psi_3-\psi_1$ \\
& $D^-$ & $D^-$ & $D^+$ & $D^+$ \\
\hline
$0$ & $-$ & $-0.477$ & $-$ & $-$ \\
$0.01$ & $-0.345$ & $-0.530$ & $0.0326$ & $-0.7307$ \\
$0.02$ & $-0.344$ & $-0.535$ & $0.0333$ & $-0.7312$ \\
$0.03$ & $-0.341$ & $-0.539$ & $0.0340$ & $-0.7316$ \\
$0.04$ & $-0.335$ & $-0.542$ & $0.0349$ & $-0.7318$ \\
$0.05$ & $-0.329$ & $-0.546$ & $0.0358$ & $-0.7320$ \\
\end{tabular}
\end{table}

\begin{figure} 
\centerline{\includegraphics[width=1\textwidth,clip=]{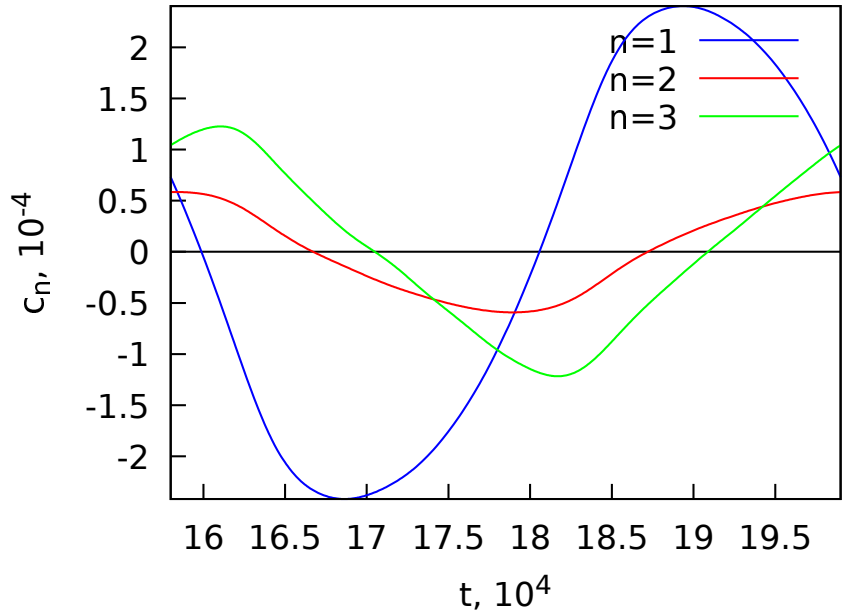}}
\caption{Coefficients $c_n(t)$ of multipole decomposition of $B_r$ at $r=r_1$ on time $t$ and $n=1,2,3$ with solar-like parameters: $D^-$, $d=0.025$, $r_1=2.85$. Dimensionless coefficients $c_n(t)$ and time $t$ are normalized to $10^{-4}$ and $10^4$ respectively.}
\label{fig:coeff_mult}
\end{figure}

The calculated phase shifts $\psi_n-\psi_1$ are practically independent of $r_1$ and weakly dependent on the asymmetry coefficient, as shown in the Table \ref{tab:phas_shft}. The simultaneous coincidence of relations $\delta_2/\delta_1$, $\delta_3/\delta_1$, and phase shift $\psi_2-\psi_1$ obtained from solar data in Equation \ref{Asun} on the basis of {\it Ulysses}' observations, are possible only for one dynamo model with $D^-$ and $d=0.025$. In this case, after selecting the value $r_1$ to ensure a coincidence with the solar values (Equation \ref{Asun}) we get $r_1=2.85$ and
\begin{equation*}
\begin{gathered}
\delta_2/\delta_1=0.23, \quad 
\delta_3/\delta_1=0.44, \\
\psi_2-\psi_1=-0.34\pi, \quad 
\psi_3-\psi_1=-0.54\pi.        
\end{gathered}
\end{equation*}
Using parameters found above ($D^-$, $d=0.025$, $r_1=2.85$) we can do the multipole decomposition in the form
$$
B_r\Bigr|_{r=r_1}=\sum\limits_{n=1}^{\infty} c_n(t) l_n(\cos\theta) 
$$
that can be obtained from Equation \ref{Br_res} with coefficients  $c_n(t)$ shown in Figure~\ref{fig:coeff_mult}. Note that dependencies  $c_n(t)$ on $t$ actually seem to be quite periodic.

\section{Discussion and Conclusions}
In this work we investigated the possible structure and dynamics of current sheets in the astrosphere. The considered dynamo model showed that for negative $D^-$ the conical current sheet should move in the opposite direction with respect to the direction of magnetic dynamo waves, which propagate equatorwards while the conical current sheet propagates polewards. For positive $D^+$ magnetic waves propagate polewards while the conical current sheet propagates equatorwards. This appears to be the most model-independent part of the results.

Our modeling also demonstrates that changing the sign of dynamo number that obtain a magnetic configuration with quadrupole symmetry rather than the initial dipole configuration. Quadrupole dynamo driven magnetic configuration are known for spherical dynamos. For more or less free changing amplitude and configuration of dynamo drivers, quadrupole configurations appear in spherical dynamo modeling more or less as usual as dipole ones \citep[e.g.][]{Metal08}.

Because the aim of our modeling was to demonstrate how rich is the variety of possible configurations, we are happy for the time being with the parameters of dynamo drivers that produce magnetic configurations of different symmetries (dipole and quadrupole ones) for negative $D^-$ and positive $D^+$.

Quite clearly the equatorial current sheet appears for dipole or octupole configurations. It seems to be a useful test to recognize quadrupole magnetic configuration in observational data.

The structure of the current sheet and relative amplitudes of various spherical harmonics of magnetic field appears to be dependent on the degree of asymmetry between the two stellar hemispheres. It is quite unexpected that the link between asymmetry and current-sheet structures looks to be substantially more pronounced for dipole configurations than for the quadrupole ones. What about the dipole case, for which structure of current sheets looks more similar to the solar one for $d=0.01$ while the relative amplitudes of dipole, quadrupole, and octupole components looks similar to the solar data obtained by {\it Ulysses} for $d=0.02$. In any case, several percent hemispherical asymmetry looks very moderate according to expectations in dynamo modeling.

We stress that if the hemispherical asymmetry of dynamo drivers is absent we obtain only odd or only even zonal magnetic-field harmonics, while the solar magnetic field does contain even and odd harmonics simultaneously. In other words, relative amplitudes of odd and even zonal magnetic harmonics give the hemispheric asymmetry of the dynamo drivers. 

Our finding that for the solar-like case current sheets propagate polewards while  dynamo wave  propagates equatorwards may appear quite a counter intuitive phenomenon. Let us briefly highlight the state of affairs in the investigation of heliospheric current-system evolution. Generally the idea of the existence of multiple evolving conical current sheets including the heliospheric current sheet remains to be investigated both theoretically and experimentally. But today we have indirect evidence that the large-scale heliospheric current sheet can not be the only one in the heliosphere. To study the evolution of the heliospheric current system, new space missions like {\it Ulysses} are needed, which could carry out measurements in the entire range of latitudes. Initially the dominant paradigm was that only one disk-like heliospheric current sheet can exist \citep[see, e.g., reviews by][]{2009SSRv..143...85B,2013BaloghSSRv..176..177B}. 

Later the question arose as to how to interpret the appearance of high-latitude current sheets in the northern and/or southern hemispheres \citep{1969Natur.222..652S, 1982SoPh...77..363L, 1996JGR...101.2475V}. Some articles suggested that the available experimental data on the intersections of the heliospheric current sheet at high latitudes should be interpreted as the presence of a second current sheet, in addition to the heliospheric one \citep{ 1969Natur.222..652S,2014ApJ...780..103W}. In the works by \cite{2001JGR...10615819S} and by \cite{2012ApJ...755..135R} on the basis of different methods ({\it Ulysses} data and  measurement of latitudinal positions of streamers in synoptic maps of the white-light corona) the important solar-wind characteristics during the solar-maximum period were shown: the radial magnetic field has clear asymmetry; the heliospheric current sheet rises up to high latitudes and there should acquire the cone shape. 

The existence of quasi-stable conical current sheets in the high-latitude heliosphere was confirmed by \cite{2017KhabarovaApJ...836..108K} based on {\it Ulysses} data, where it was shown that such the conical current sheet exists and has vortex-like structure of current. Two-sector latitudinal extent of the distribution of heliospheric current sheet during solar maxima from 1976 to 1994 was mentioned by \cite{1996JGR...101.2475V} with help of the source-surface model. Also investigations of scattered Lyman-$\alpha$ radiation in heliosphere allowed the reconstruction of the dependence of the solar-wind mass flux on helio-latitude \citep{2019KatushkinaSoPh..294...17K}. It was shown that at solar minima the dependence of the concentration on helio-latitude corresponds to the usual position in the region of the Sun's Equator and minima at the Poles. However, two concentration maxima are clearly seen at mid-latitudes near solar maximum \citep{2020PhyU...63..801P}. Therefore the indication of the existence of two high-latitude current sheets during solar maxima periods is present. 

\begin{figure} 
\centerline{\includegraphics[width=1\textwidth,clip=]{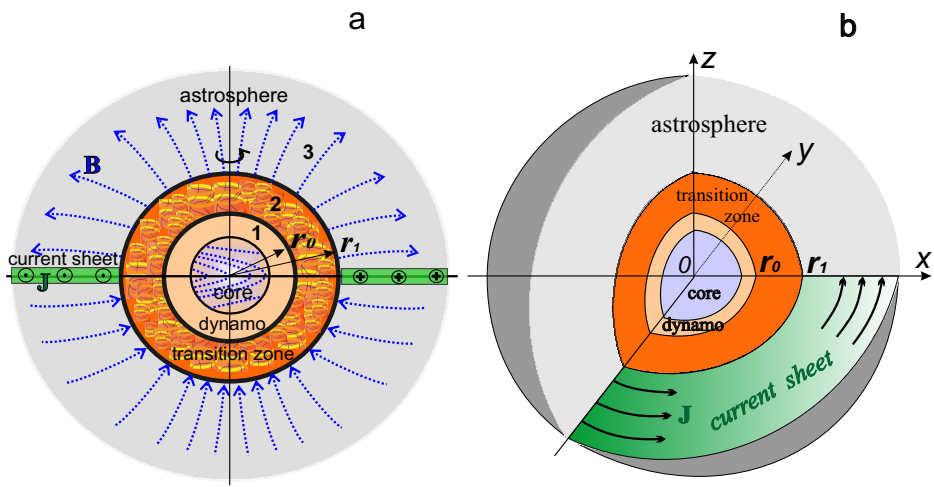}}
\caption{Schematic representation of the modeled spherical zones where magnetic field  is generated in a spherical dynamo zone 1, then transformed in the transition zone 2; finally it spreads outward into the astrosphere (external zone 3) in the form of a large-scale magnetic field: (\textbf{a}) the general view of the structure of corresponding magneto-active zones with large-scale current sheet with current density $J$; (\textbf{b}) the corresponding 3D view of disk-like current sheet and the section of magneto-active zones, that were taken into account in the model. At the left the internal \textquotedblleft core\textquotedblright zone is marked but it is surrounded by  spherical zone with a winding magnetic lines involved in the dynamo processes.
}\label{fig:magnetosph}
\end{figure}

According to current views, the heliospheric current sheet (i.e. magnetic neutral surface) is a continuation of the neutral line in the corona of the Sun. During the solar-activity cycle the neutral line changes its topology as well as latitudinal position \citep{2008ApJZhukov...680.1532Z}, consequently the shape and position of the heliospheric current layer should also change \citep{2014WangSSRv..186..387W,2000JGRWang...10525133W}. The problem of the formation and evolution of multiple conical current sheets in the heliosphere corresponding to the cycle of the solar activity were investigated and discussed in details by \cite{maiewski2020magnetohydrodynamic}. It was shown that conical current sheets in the heliosphere should move in the northern direction, corresponding to the motion of neutral lines of the multipole composition of the helio-magnetic field. Therefore it is characteristic for the Sun that the activity wave in the main bulk of the solar surface propagates equatorwards (as we mentioned above) while the magnetic neutral line migrates poleward \citep{Metal83}. The corresponding evolution and spatial motion of conical current sheets in the heliosphere are illustrated  by Figure~\ref{fig:magnetosph} and described in Section 6.1.

Note that here we consider dynamo drivers as a given quantity while in reality they are determined by stellar convection and rotation. Of course, it is done just to isolate the statement of the problem under discussion. Nevertheless the degree of hemispheric asymmetry of dynamo drivers obtained to explain solar quadrupole moment is comfortably small enough to be explained just by pure statistical noise arising for averages taken over a not very large ensemble of convective cells, see e.g. \cite{KK21}. A more delicate point is consideration of poleward-propagating activity waves. Of course, solar-dynamo studies investigate why distribution solar-dynamo drivers produce the equatorward propagating activity wave. It would be however an exaggeration to insist that solar experiences are fully instructive for stellar-rotation laws even for binary stellar systems or even for exoplanetary systems with hot super-Jupiters close to the host star. Even less definite is our knowledge concerning distribution of mirror asymmetry or meridional circulation in various stars. In our opinion it looks reasonable to consider the poleward-propagating activity waves at least until we learn that observations reject this option and confirm that the solar example remains instructive for various stellar systems.

It should be noted that in this model we used the simplest system of three spherical regions of the magnetic field, where the main changes of the stellar magnetic field are taken into account from its formation in the dynamo region to its transformation into a large-scale magnetic field that is carried out in the radial direction from the star's corona outward to the astrosphere. Figure~\ref{fig:magnetosph} illustrates schematically the location of zones of magnetic-field changes that might be unique for solar-like and the other stars having proper magnetic fields. Thus Figure~\ref{fig:magnetosph}a shows the process of generating a poloidal (dipole) magnetic field from a toroidal one in a thin spherical dynamo region (number 1). Its passage through the transition region 2 leads to the formation of a large-scale dipole magnetic field in zone 3. Like the solar magnetic field, it propagates along with the flow of magnetized plasma outward to the astrosphere and spreads over vast distances (the size of heliosphere is about 100\,AU). While usually in the theoretical works \citep[e.g.][]{1993UsmanovSoPh..143..345U, 2014UsmanovApJ...788...43U} the observable shapes of the neutral lines in the Sun's corona are taken into account, in our composite model only the general characteristics of large-scale magnetic fields of stars are important for consideration. We try to understand the fundamental interaction between dynamo processes in the interior and the processes that lead to formation of a large-scale disk-like or conical current sheets in the astrospheres. Thus, one can see in Figure~\ref{fig:magnetosph}a,b two main boundaries separating zones 1 and 2 and zones 2 and 3. At the first boundary, we define the solution obtained from the dynamo equations and use some reasonable simplifications. At the second boundary we use the solution for magnetic-field transformation described above. As a result of this transformation, the external magnetic field should acquire substantially new properties in comparison with the interior, i.e. magnetic lines become predominantly radial, while such structures as multiple coronal magnetic loops practically disappear in a wide range of latitudes.

Our results showed that the structure of heliospheric and astrospheric current sheets and, consequently, the entire astrospheres, depends on the signs and values of the dynamo numbers characterizing the dynamo model. This indicates that the internal-dynamo processes in the interior of the Sun or Sun-like stars can determine the structure of their astrospheres \citep[e.g.][]{1999SoPh..188..277E, 2015ApJ...798..116R, maiewski2020magnetohydrodynamic}, and a variety of large-scale current systems in astrospheres significantly goes beyond our knowledge about dynamo processes in the solar system. It is shown that the magnetic fields of the neutral lines in the upper layers of the solar or stellar coronas can be extended outwards where they are smoothed out, averaged, then filled by plasma, forming evolving disc-like or conic-like current sheets. After reaching the Alfv\'en surface, these current sheets propagate in almost unchanged state over the whole astrosphere, therefore determining the peculiarities of its large scale structure. As a result, the formation of a large-scale current system takes place in accordance with the dominating kind of the stellar multipole magnetic field and its symmetric or asymmetric distribution in the northern and southern hemispheres of the stars \citep{sym12122085, maiewski2020magnetohydrodynamic}. The evolution of stellar current sheets may also be fundamentally different from what we observe in the solar system.

\begin{acks}
E.Maiewski, H.Malova, V.Popov, D.Sokoloff acknowledge the support of Ministry of Science and Higher Education of the Russian Federation under the grant №075-15-2020-780 (N13.1902.21.0039). The {\it Ulysses} Final Archive is available at the website \href{http://ufa.esac.esa.int/ufa/#data}{http://ufa.esac.esa.int/ufa/data} where the directory VHM-FGM contains the {\it Ulysses} prime resolution data of the magnetic field. The {\it Ulysses} Orbital Information (1990\,--\,2009) is available as follows: \href{https://www.cosmos.esa.int/web/ulysses/orbit}{https://www.cosmos.esa.int/web/ulysses/orbit}. E.Yushkov performed dynamo simulations with the financial support of the Ministry of Education and Science of the Russian Federation as part of the program of the Moscow Center for Fundamental and Applied Mathematics under the agreement №075-15-2019-162.
\end{acks}

\begin{ethics}
\begin{conflict}
The authors declare that they have no conflicts of interest.
\end{conflict}
\end{ethics}

\bibliographystyle{spr-mp-sola}
\bibliography{2022_SolarPhysics_Sokoloff}  

\end{article} 
\end{document}